\documentclass[conference]{IEEEtran}
\IEEEoverridecommandlockouts
\usepackage{microtype} 
\usepackage{cite}
\usepackage{amsmath,amssymb,amsfonts}
\usepackage{algorithmic}
\usepackage{graphicx}
\usepackage{textcomp}
\usepackage{threeparttable}
\usepackage{xcolor}
\def\BibTeX{{\rm B\kern-.05em{\sc i\kern-.025em b}\kern-.08em
    T\kern-.1667em\lower.7ex\hbox{E}\kern-.125emX}}
\begin{document}

\title{FastAudio: A Learnable Audio Front-End for Spoof Speech Detection\\
}
\author{\IEEEauthorblockN{Quchen Fu, Zhongwei Teng, Jules White, Maria Powell, and Douglas C. Schmidt}}

\maketitle

\begin{abstract}
Voice assistants, such as smart speakers, have exploded in popularity. It is currently estimated that the smart speaker adoption rate has exceeded 35\% in the US adult population. Manufacturers have integrated speaker identification technology, which attempts to determine the identity of the person speaking, to provide personalized services to different members of the same family. Speaker identification can also play an important role in controlling how the smart speaker is used. For example, it is not critical to correctly identify the user when playing music. However, when reading the user's email out loud, it is critical to correctly verify the speaker that making the request is the authorized user. Speaker verification systems, which authenticate the speaker identity, are therefore needed as a gatekeeper to protect against various spoofing attacks that aim to impersonate the enrolled user. This paper compares popular learnable front-ends which learn the representations of audio by joint training with downstream tasks (End-to-End). We categorize the front-ends by defining two generic architectures and then analyze the filtering stages of both types in terms of learning constraints. We propose replacing fixed filterbanks with a learnable layer that can better adapt to anti-spoofing tasks. The proposed FastAudio front-end is then tested with two popular back-ends to measure the performance on the LA track of the ASVspoof 2019 dataset. The FastAudio front-end achieves a relative improvement of 27\% when compared with fixed front-ends, outperforming all other learnable front-ends on this task.

\end{abstract}


\section{Introduction}
The prevalence of smart appliances has streamlined our daily chores but also brought security concerns. It is currently estimated that over 35\% of the US adult population has a smart speaker at home\cite{Bret2021USsmart}. These voice assistants are becoming ever more versatile and can automate tasks ranging from making a phone call to placing an order. However, many of these tasks require different levels of privileges that are tied to the identity of the person talking to the voice assistant. Therefore, identifying “who is speaking” has become the backbone of personalized voice services. Compared to Speaker Identification, which focuses on personalized services, Speaker Verification is a binary classification of the validity of user identity claims for biometric security. 

Spoofing attacks against speaker verification approaches exist and can be broadly categorized into three groups: Text-To-Speech (TTS), Voice Conversion (VC), and Replay Attacks. Replay attacks, which attack the identification system by playing back the recorded sample of the victim's speech, are most prevalent as they require the least technological sophistication. However, the threat of replay attacks can be mitigated by adding random prompt words. With the rapid development of deep learning, TTS and VC have seen significant improvement in their ability to fool speaker verification systems. Popular models like the Tacotron2\cite{Shen2018NaturalTS} can transform text into high-quality synthetic speech that is almost indistinguishable from the speech of humans.

Audio files are usually stored as 1D vectors that are extremely long (1 second of audio recording with a sampling rate at 16kHz contains 16000 data points). Because of their length, they are traditionally prepossessed to create a compressed representation that is smaller in size but aims to preserve as many of the important features as possible before spoof detection is applied. The component that performs this preprocessing step is known as the front-end. Front-ends can be either handcrafted or learnable, and the process of choosing the proper handcrafted front-ends is also known as feature selection. Both types of front-ends contain filter layers, and constraints can be applied to the filters.

Though handcrafted front-ends have proven to be a strong baseline for a variety of tasks, the underlying idea guiding the design of these features is that they are modeled on the non-linearity of the human ear's sensitivity to frequency (Mel scale) and loudness (Log compression). 

Therefore, they may not represent the most salient features for audio classifications under all domains. Empirically, learnable front-ends outperform handcrafted front-ends in 7 out of 8 audio classification tasks in recent studies\cite{Zeghidour2021LEAFAL}.

This paper provides the following contributions to the study of defending against audio spoofing attacks:

\begin{enumerate}

    \item It proposes a light-weight\footnote{Front-end trains faster and has the least computational complexity as estimated by multiply–accumulate operations (MACs) compared to other learnable front-ends. See Table \ref{table:frontend_compare}, https://pypi.org/project/ptflops/.} learnable front-end called FastAudio that achieved the lowest min t-DCF in spoof speech detection compared to other front-ends,
    \item It provides a comparison of feature selections for spoofing countermeasures, with a special focus on learnable audio front-ends, and shows how applying shape constraints can make the filterbank layer perform better while reducing the number of parameters, and
    \item It describes the architecture that achieved top performance on the ASVspoof 2019\cite{Todisco2019ASVspoof2F} dataset.
    
\end{enumerate}

The remainder of this paper is organized as follows: Section~\ref{bg.section} summarizes the classification of audio front-ends based on structure and the background for filter learning; Section \ref{question.section} discusses different constraint types regarding filterbank learning; Section \ref{backend.section} describes our experiment setups including dataset, metric, and model details;
Section \ref{results.section} analyzes the result and describes our experiment insights regarding filter learning for spoof speech detection, and Section \ref{conclusionfuturework.section} presents concluding remarks and potential future work.

\section{Background on Audio Front-end}
\label{bg.section}
Spoof speech detection is a single-task classification problem, for which many front-ends have been tested, including Instantaneous Frequency (IF), Group Delay (GD) and Mel-frequency cepstral coefficients (MFCC), etc. The front-ends used in the classification of speech have been dominated by MFCC and recently Log Mel FilterBanks (FBanks); both are hand-crafted features that are fixed and not learnable. Constant Q Transform (CQT)\cite{Brown1991CalculationOA} is another handcrafted front-end commonly used for music generation and music note recognition as it can better mimic musical scales; however, prior research reported CQT was also the best performing front-end for spoof detection\cite{Li2021ReplayAS}. 

\begin{figure}[hpbt]
\centering
\includegraphics[width=1.0\linewidth]{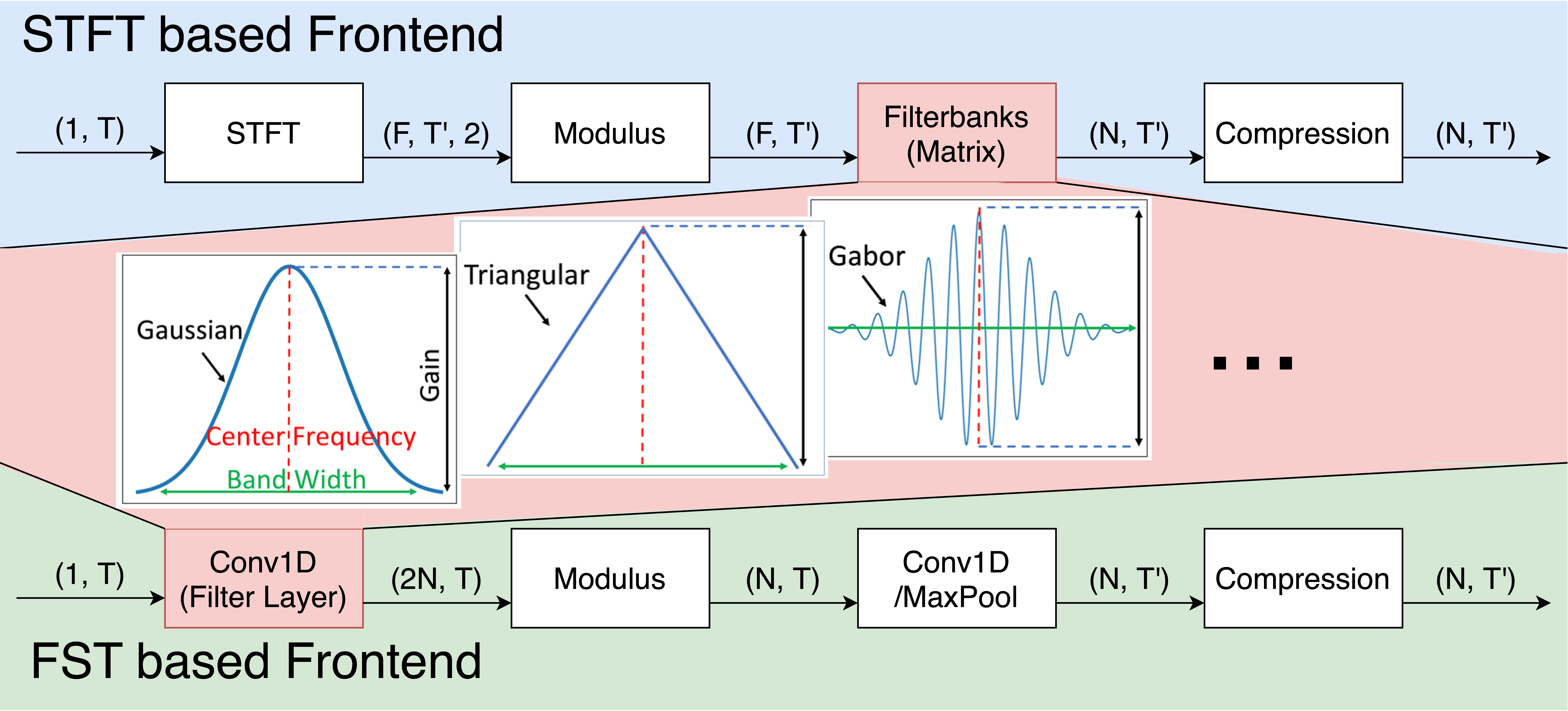}
\caption{FST based front-end and STFT based front-end}
\label{fig:fst_stft}
\end{figure}

As shown in Figure \ref{fig:fst_stft}, front-ends can be categorized by the procedures they perform. There are two key categories: First-order Scattering Transform (FST)\cite{Andn2014DeepSS} based front-ends and Short-Time Fourier Transform (STFT) based front-ends. Unlike STFT which multiplies a filterbank matrix with a spectrogram,  FST uses a convolutional layer on the raw audio waveform to approximate the standard filtering process. While considerable progress has been made on FST based front-end approaches, literature has shown that they lose signal energy, which corresponds to information loss, since only the first-order coefficients of a scattering transform are used\cite{Andn2014DeepSS}. The FST based approaches are also time-consuming\cite{Lee2018SampleCNNED} since convolution layers with large kernels are computation intensive. STFT based front-ends remain popular, and FBanks are still the front-end for the state-of-the-art speaker identification\cite{Desplanques2020ECAPATDNNEC} and speech recognition\cite{Villalba2020StateoftheartSR} systems. However, many STFT based front-ends are fixed and may not adapt well to certain downstream tasks.

Both types of front-ends employ some type of filter-like manipulations to model the non-linearity of the human ear's sensitivity to frequency. The distribution of filter center frequency is referred to as scale. Studies~\cite{Lippmann1997SpeechRB} have shown that the Mel-scale, as shown in Equation~\ref{eq:mel}, can capture human perception for pitch relatively well. 

\begin{equation}
m=2595 \log _{10}\left(1+\frac{f}{700}\right) \label{eq:mel}
\end{equation}

There also exists the Bark scale~\cite{Zwicker1980AnalyticalEF} and Equivalent Rectangular Bandwidth (ERB) scale~\cite{Glasberg1990DerivationOA}, which are less well-known. However, these scales are mostly based on past experience and are fixed equations. To make this manipulation in the front-end domain adaptable, filters can be made learnable. As shown in Figure~\ref{fig:fst_stft}, a filterbank can learn its center frequency $c_n$, gain $g_n$, bandwidth $b_n$, and shape $s_n$. The filter properties can be summarized in Equation \ref{eq:filter}.

\begin{equation}
\omega_n(f)=g_ns_n(c_n; b_n; f) \label{eq:filter}
\end{equation}

Recent research has made progress on learnable audio front-ends. SincNet\cite{Ravanelli2018SpeakerRF} uses convolution to extract features with Sinc functions. TD-filterbanks\cite{Zeghidour2018LearningFF} uses Gabor convolution to replace front-end filtering. LEAF\cite{Zeghidour2021LEAFAL}, proposed by Google, is the first fully learnable audio front-end with an added learnable compression layer. Wav2Vec\cite{Baevski2020wav2vec2A} is a CNN-based unsupervised audio training method for speech recognition that can use raw audio data directly for training. RawNet2\cite{Tak2021EndtoEndAW} was used as the baseline system for the ASVspoof 2021 challenge and adopted Sinc Filters with CNN layers to extract audio features.

\section{Research Questions Regarding Learnable Filterbanks}
\label{question.section}
Previous research has explored the feasibility of learnable filterbanks. For example, nnAudio\cite{Cheuk2020nnAudioAO} implemented a set of unconstrained learnable filterbanks; however, T. Sainath \textit{et al.}\cite{Sainath2013LearningFB} reported limited improvement from unconstrained filterbank learning. DNN-FBCC\cite{Yu2017DNNFB} explored some constraints over filters by adopting a mask matrix. Zhang and Wu\cite{Zhang2019DiscriminativeFF} described a detailed study on the shape and positiveness constraint's effect on the filterbanks. However, no systematic study has been done on constraining the filterbank shape in the STFT-based approach used for spoof speech detection. 

\begin{table}[htbp]
\caption{\label{table:frontend-filter}Filter Comparison of Learnable Front-end}
\centering
  \begin{threeparttable}
\resizebox{\columnwidth}{!}{
\begin{tabular}{c|c|c|c|c|c|c}
\hline
  \textbf {Type} & \textbf{Name}  & \multicolumn{2}{c|}{\textbf{Filter/BandWidth}} & \multicolumn{2}{c|}{\textbf{Center Frequency}} & \textbf{Gain}\\
\cline{3-6} 
 & & \textbf{Shape} & \textbf{Clamp} & \textbf{Sorted} & \textbf{Clamp} \\
\hline
\textbf{FST} & TD-FBanks & Gabor & - & Yes & Yes & -\\
\cline{2-7} 
\textbf{based}  & SincNet & Sinc & Yes &  No & Yes & Fixed\tnote{a}\\
\cline{2-7} 
\textbf & LEAF & Gabor & Yes & No & Yes & Fixed\\
\hline
\textbf{STFT} & nnAudio & - & No & No & No & -\\
\cline{2-7} 
\textbf{based}  & DNN-FBCC & - & Yes & Yes & Yes & -\\
\cline{2-7} 
\textbf & FastAudio & Triang & Yes & No & Yes & Fixed\\
\hline
\end{tabular}}
\begin{tablenotes}
    \item[a] The gain of each filter is not learned in the filter layer but in subsequent layers.  
\end{tablenotes}
  \end{threeparttable}
\end{table}

\begin{table*}[hbt!]
\caption{\label{table:stftfrontend}A stage-wise comparison of STFT based front-end}
\begin{center}
\begin{tabular}{|c|c|c|c|c|c|c|c|c|}
\hline
\textbf {Type} & \textbf{Name} & \textbf{Pre-emph} & \textbf{FFT} & \textbf{Selection} & \textbf{Filter} & \textbf{Compression} &\textbf{Transform} & \textbf{Center freq}\\
\hline
\textbf  & Spectrogram & - & STFT & Modulus & - & Log (optional) & - &-\\
\cline{2-9} 
\textbf  & Gammatonegram &  - & STFT & Modulus & Gammatone & Log (optional) & - &Mel\\
\cline{2-9} 
\textbf  & IF Derivative &  - & STFT & Phase & - & - & Derivative & -\\
\cline{2-9} 
\textbf  & IF &  - & STFT & Phase & - & - & - & -\\
\cline{2-9} 
\textbf {Fixed} & Mel-Filterbanks &  - & STFT & Modulus & Triangular & Log & - & Mel\\
\cline{2-9} 
\textbf {Hand-crafted} & MFCC &  Yes & STFT & Modulus & Triangular & Log & DCT & Mel\\
\cline{2-9} 
\textbf {Feature} & LFCC &  - & STFT & Modulus & Triangular &  Log & DCT & Linear\\
\cline{2-9} 
\textbf {Extractor} & IMFCC &  - & STFT & Modulus & Triangular &  Log & DCT & Inverse Mel\\
\cline{2-9} 
 & RFCC &  - & STFT & Modulus & Rectangular &  Log & DCT & Linear\\
\cline{2-9} 
 & GFCC &  - & STFT & Modulus & Gammatone &  Log & DCT & ERB\\
\cline{2-9} 
\textbf  & CQT &  - & STFT & Modulus & Constant-Q & Log (optional) & - & CQT\\
\cline{2-9} 
\textbf  & CQCC &  - & STFT & Modulus & Constant-Q & Log & DCT & CQT\\
\cline{2-9} 
\textbf  & SPNCC/PNCC &  Yes  & STFT & Modulus & Gammatone & Log & DCT & ERB \\
\cline{2-9} 
\textbf  & RASTA-PLP &  -  & STFT & Modulus & RASTA & Log & IDFT & RASTA*\\
\hline
\textbf {Partly} & PCEN &  -  & STFT & Modulus & Triangular & PCEN & - & Mel\\
\cline{2-9} 
\textbf {Trainable} & Spline &  -  & STFT & Modulus & Spline & PCEN & - & Mel\\
\cline{2-9} 
\textbf  & nnAudio &  -  & Trainable STFT & Modulus & Conv1D & Log & - & Mel/CQT\\
\cline{2-9} 
\textbf  & DNN-FBCC &  -  & STFT & Modulus & Matrix Mask & Log & - & Mel\\
\cline{2-9} 
 & \textbf{FastAudio} &  Yes & \textbf{STFT} & \textbf{Modulus} & \textbf{Triangular} & \textbf{Log} & - & \textbf{Mel(Trainable)}\\
\hline
\end{tabular}
\end{center}
\end{table*}

\begin{table*}[htb!]
\caption{\label{table:DSPfrontend}A stage-wise comparison of Deep Scattering Spectrum based front-end}
\begin{center}
\begin{tabular}{|c|c|c|c|c|c|c|c|}
\hline
\textbf {Type} & \textbf{Name} & \textbf{Pre-emph} & \textbf{Filter} & \textbf{Selection} & \textbf{Windowing/Pooling} & \textbf{Compression}  & \textbf{Initial Center freq}\\
\hline
\textbf {Partly} & TD-FBanks & - & Conv1D & Modulus & Lowpass& Log & Mel(Trainable)\\
\cline{2-8} 
\textbf {Trainable} & SincNet & - & Sinc & LeakyRelu& Maxpool & LayerNorm & Mel\\
\hline
\textbf {Trainable} & LEAF & - & Gabor & Modulus & Gaussian Lowpass & sPCEN & Mel(Trainable)\\
\cline{2-8} 

\hline
\end{tabular}
\label{tab1}
\end{center}
\end{table*}

As shown in Table \ref{table:frontend-filter}, all current FST based front-ends put shape constraints on the band-pass filters; however, STFT based front-ends, like DNN-FBCC, do not constrain the filter shape. Instead, a mask is put on the filters so that the bandwidth is clamped and the filters are sorted by center frequencies. Therefore, we designed a learnable front-end, called FastAudio, specifically focused on answering the following questions: 

\begin{enumerate}
    \item Is a shape constraint necessary for spoof detection, and which shape constraint has the lowest min t-DCF? 
    \item Does the center frequency need to be sorted for spoof detection? 
    \item What do trained filterbanks learn about spoof detection compared to handcrafted FBanks?
\end{enumerate}
These questions are discussed in subsection \textit{C} and \textit{D} of Section \ref{results.section}.

\section{Experiment and Dataset}
\label{backend.section}
The ASVspoof 2019 corpus consists of two parts: Logical Access and Physical Access. Logical Access (LA) contains fake (spoof) speech generated from various text-to-speech and voice conversion techniques. Physical Access (PA) contains spoof speech that is simulated to mimic various room sizes, speaker orientations/distances, and hardware artifacts. The true speech audio files are referred to as \textit{Bona fide}. Here we focus on the LA task. Since there are existing ASV (automatic speech verification) systems that provide some protection against spoofing attacks, the goal is to design a system that can best complement existing ASV systems (the result of the existing ASV system is provided by the dataset in labels). The system we are designing is called the countermeasures (CM). The evaluation metric is the tandem detection cost function (min t-DCF), which is designed to best reflect real-world protection effects.

\subsection{Dataset}

The performance of the FastAudio learnable front-end is evaluated on the ASVspoof 2019 LA dataset. As shown in Table \ref{table:dataset}, the dataset was partitioned into three parts where the evaluation set is three times the size of the training set. The training and development sets contain data generated from the same algorithms; however, to ensure the spoof detection system can generalize well to audio of unseen types, the evaluation set also contains attacks that are generated from different algorithms.
\begin{table}[htbp]
\caption{\label{table:dataset}DESCRIPTION OF ASVSPOOF 2019 LA DATASET}
\begin{center}
\begin{tabular}{c|c|c|c|c}

\hline \textbf{Subset} & \multicolumn{2}{|c|} { \textbf{\#Speaker} } & \multicolumn{2}{c} { \textbf{\#Utterances} } \\
\cline { 2 - 5 } & \textbf{Male} & \textbf{Female} & \textbf{Bona fide} & \textbf{Spoofed} \\
\hline Training & 8 & 12 & 2580 & 22800 \\
\hline
Development & 8 & 12 & 2548 & 22296 \\
\hline
Evaluation & 21 & 27 & 7355 & 63882 \\
\hline
\end{tabular}
\end{center}
\end{table}

\begin{table*}[htb!]
\caption{\label{table:architecture} A stage-wise comparison of different Architectures}
\begin{center}
\begin{tabular}{c|c|c|c|c|c|c}
\hline
\textbf {Name} & \textbf{Front-end} & \textbf{First Layer} & \textbf{Main Block} & \textbf{Pooling} & \textbf{Optional} &\textbf{Classifier} \\
\hline
\textbf {X-vector} & MFCC & TDNN & CNN & Statistical Pooling & - & PLDA \\
\hline
\textbf {ECAPA-TDNN} & FBank & TDNN & CNN & Attention Statistical Pooling & PLDA & Cosine Similarity\\
\hline
\textbf {Res2Net} & CQT & TDNN & CNN & Pooling & PLDA & Cosine \\
\hline
\textbf {RawNet2} & SincNet & Conv & CNN & GRU & - & FC \\
\hline
\textbf {EfficientNetB0} & LEAF & Conv & CNN & Avg/Max pooling & - & FC \\
\hline
\end{tabular}
\end{center}
\end{table*}

\subsection{Metrics}

The primary metric for spoof speech detection is the minimum normalized tandem detection cost function (min t-DCF), as shown in Equation \ref{eq:TDCF}. The min t-DCF measures the overall protection rate for combined CM and ASV systems, where $\beta$ depends on application parameters (priors, costs) and ASV performance (miss, false alarm, and spoof miss rates), while $P_{\mathrm{miss}}^{\mathrm{cm}}(s)$ and $P_{\mathrm{fa}}^{\mathrm{cm}}(s)$ are the CM miss and false alarm rates at threshold $s$\cite{Todisco2019ASVspoof2F}.

\begin{equation}
\mathrm{t-DCF}_{\mathrm{norm}}^{\min }=\min _{s}\left\{\beta P_{\mathrm{miss}}^{\mathrm{cm}}(s)+P_{\mathrm{fa}}^{\mathrm{cm}}(s)\right\}\label{eq:TDCF}
\end{equation}

Equal error rate (EER) was used as a secondary metric to make comparison possible with earlier datasets like ASVSpoof 2017. EER is defined as the value of false acceptance rate and false rejection rates where they are equal.

\subsection{Back-end}
We made a summary on the structures of common front-ends in Table \ref{table:stftfrontend} and Table \ref{table:DSPfrontend}. Our FastAudio front-end consists of an STFT transform followed by a learnable filterbank layer, and finally a log compression layer to mimic the non-linearity of human sensitivity to loudness. We integrated the front-ends with two of the most popular back-ends for audio classification: X-vector\cite{snyder2018x}\cite{Ravanelli2021SpeechBrainAG} and ECAPA-TDNN\cite{desplanques2020ecapa}\cite{Ravanelli2021SpeechBrainAG}. The back-end turns the FBank-variant into a 256-dimensional embedding vector. The vectors are then fed into a linear classifier. A summary of other popular back-end architectures is shown in Table \ref{table:architecture}.

\begin{table}[bhtp]
    \caption{\label{table:backend} X-vector and ECAPA-TDNN}
\centering
\begin{tabular}{ll|ll}
\hline 
\multicolumn{2}{c|}{\textbf{Xvector}} & \multicolumn{2}{c}{\textbf{ECAPA-TDNN}} \\
\hline 
\textbf{Layer} & \textbf{Output} &  \textbf{Layer} & \textbf{Output} 
\\\hline 
Input & (N, T${'}$) & Input & (N, T${'}$) 
\\\hline 
TDNN X 5  & (1500, T${'}$) & Conv1D + ReLU + BN & (C, T${'}$)
\\\hline 
Stats Pool  & (3000, 1) & SE-Res2Block X 3 & (3, C, T${'}$)
\\\hline 
Linear &  (256, 1) & Conv1D + ReLU & (1536, T${'}$)
\\\hline 
& & Atten Stats Pool + BN & (3072, 1)
\\\hline 
& &  FC + BN & (256, 1)
\\\hline 
\end{tabular}
\end{table}

\subsection{Experimental Setup}
The model was trained on 2 Nvidia 2080 Ti GPUs for 100 epochs and the batch size was set to 12 (except for TD-filterbank whose batch size was 4 to stay within memory limits). We also compared the performance of our front-end with other STFT-based and FST-based front-ends, both under learnable and fixed settings. To make the comparison fair, we keep the hyperparameters across all experiments the same so that the front-end outputs have the same dimensions. The sampling rate was set to 16kHz, window length to 25ms, window stride to 10ms, and the number of filters to 40. All learnable front-ends were initialized to mimic Mel-FBanks, as previous research\cite{Zeghidour2018LearningFF} has shown that random initialization has worse performance.

\section{Results and Analysis}
\label{results.section}

\subsection{\textbf{How do learnable front-ends perform on min t-DCF compared with handcrafted front-ends for spoof speech detection?}}

Since the most recent systematic comparison of front-ends on spoof detection we can find was done in 2015\cite{Xiao2015SpoofingSD}, we designed the experiments so that an updated baseline can be established that includes learnable front-ends. We choose a combination of FST and STFT front-ends with both fixed and learnable setting so that the experiment is comprehensive. As shown in Table \ref{table:frontend_compare}, we found that the FST-based learnable front-ends need longer training time than hand-crafted features in the spoof speech detection task and cannot beat the performance of CQT.

\begin{figure}[hbtp]
\centering
\includegraphics[width=1.0\linewidth]{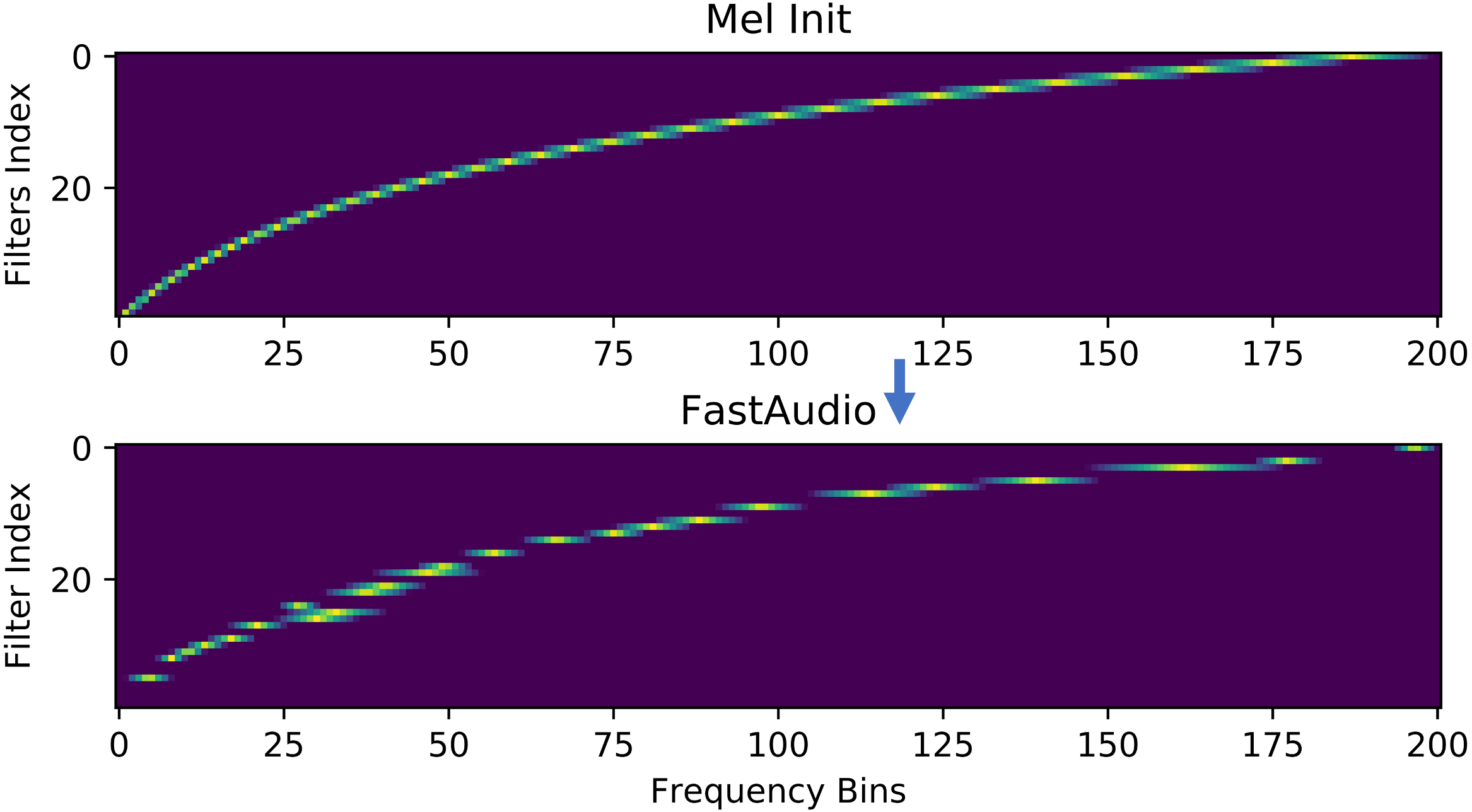}
\caption{Heatmap of the magnitude of the frequency response for initialization filters (up) and learned filters (down).}
\label{fig:fastaudio-mel}
\end{figure}

\begin{table*}[htbp]
\begin{center}
\caption{\label{table:frontend_compare} A stage-wise comparison of the different Front-ends' performance on the ASVspoof 2019 LA dataset}
  \begin{tabular}{ccccccccc}
  \hline
     & &  & \multicolumn{2}{c}{\textbf{ECAPA-TDNN}} & \multicolumn{2}{c}{\textbf{X-vector}} \\ 
    \hline
    \textbf{Front-end} & \textbf{\#Params} & \textbf{Constraint} & \textbf{EER} & \textbf{min t-DCF}& \textbf{EER} & \textbf{min t-DCF} & \textbf{MACs} & \textbf{Train Time/Epoch} \\ 
    \hline
    CQT & 0 & Fixed & 1.73 &  0.05077 & 3.40 & 0.09510 & 0 & 10:58 min\\
    Fbanks & 0 & Fixed &  2.11 &  0.06425 & 2.39 & 0.06875& 0 & 10:53 min\\
    \textbf{FastAudio-Tri} & \textbf{80} & \textbf{Shape+Clamp} &  \textbf{1.54} &  \textbf{0.04514} &  \textbf{1.73}& \textbf{0.04909}& \textbf{0.00GMac} & \textbf{13:02 min}\\
    FastAudio-Gauss & 80 & Shape+Clamp &  1.63 &  0.04710&  1.67& 0.05158& 0 & 12:51 min\\
    FastAudio-Sort & 80 & Shape+Clamp+Order & 1.89 & 0.05204 &  1.69& 0.05235& 0 & 12:59 min\\
    LEAF & 282 & Shape+Clamp & 2.49 & 0.06445 &  3.28& 0.07319& 0.01GMac & 34.45 min\\
    nnAudio &8.04k & No & 3.63 & 0.08929& 5.56 & 0.14707 & 0 & 13:00 min\\
    TD-filterbanks & 31k & Shape+Clamp&  1.83&  0.05284& 3.18& 0.08427& 1.32GMac & 22.48 min\\
    \hline
    \hline
    \textbf{Front-end} & \textbf{Name} & \textbf{Constraint} & \textbf{EER} & \textbf{min-tDCF}& \multicolumn{2}{c}{\textbf{Backend}} &  &   \textbf{Baseline}\\ 
    \hline
    SincNet & RawNet2 & Fixed &  5.13&  0.1175&   \multicolumn{2}{c}{-} &  &  - \\
    \hline
    
  \end{tabular}
 \end{center}
\end{table*}

\begin{figure*}[hpbt]
\centering
\includegraphics[width=1.0\linewidth]{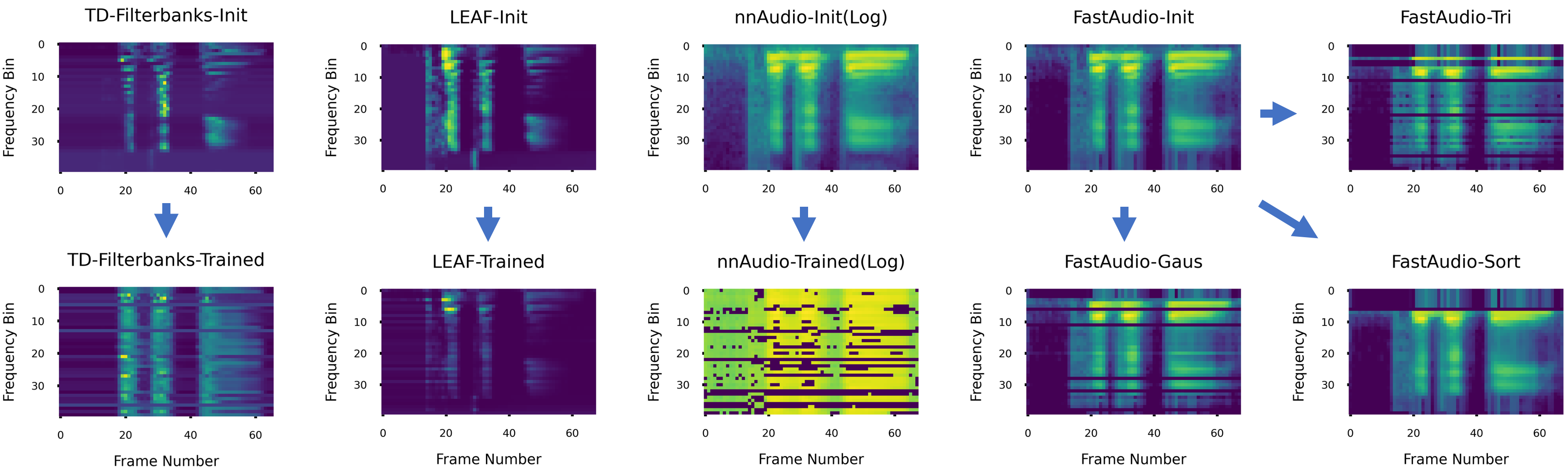}
\caption{Visualization of Learnable Front-ends}
\label{fig:learnable}
\end{figure*}

\subsection{\label{constraint.subsection}\textbf{Can we design an STFT-based front-end for spoof speech detection that is learnable and can it beat the performance of CQT?} }

Since FST-based learnable front-ends failed to beat the performance of CQT, we designed a front-end following the traditional STFT-based approach and limited the number of trainable parameters. Since it trains faster than FST-based front-ends, we call it FastAudio. We hypothesize that instead of changing the front-end architecture completely like in the FST-based approach, we can boost the performance of the fixed STFT-based approach by making the filterbank layer learnable. We tested FastAudio under 3 different constraint settings and the best one achieved 27\% decrease in min-tDCF compared to FBanks, outperforming CQT (See Table \ref{table:frontend_compare}).

\subsection{\textbf{Which set of constraints for filterbank learning performs best in spoof speech detection?}}

We found that the existence of shape constraint plays an important role in improving spoof detection accuracy. However, we did not find a significant difference in constraining the shape of the filters to be Gaussian or Triangular. We found that sorting the filterbanks by center frequency does not improve accuracy, which confirms the conclusion from previous study in LEAF\cite{Zeghidour2021LEAFAL}. As shown in Figure \ref{fig:fastaudio-mel}, the learned filterbank distribution closely follows the hand-crafted filterbanks in both center frequency and bandwidth. The similarity in $c_n$ and $b_n$ helps explain the strong performance of handcrafted features compared to the learnable front-end, especially compared to the FST-based front-ends. We hypothesize that during the training process, the filterbanks actually 'self-regulate' to remain mostly sorted in both center frequency and bandwidth. 

The visualization of the front-end output is shown in Figure \ref{fig:learnable}. All of the outputs contain "horizontal lines" that correspond to certain frequencies, which is a sign of filter selectiveness. We found that front-end output like LEAF, TD-filterbanks and nnAudio changed greatly after training due to the number of trainable parameters. When the shape of the filters is not constrained, as shown in nnAudio, the trained front-end shows signs of over-fitting (many random distributed dots) and has the worst performance. Since the nnAudio has no constraint for filter shape, the learned filter shape is determined by 201 points, thus it may contain very sharp peaks and select frequencies of very narrow ranges, thus creating the irregular dots.
\subsection{\textbf{What does FastAudio learn about spoof speech detection and how can we interpret what it learns?}}

\begin{figure}[hpbt]
\centering
\includegraphics[width=1.0\linewidth]{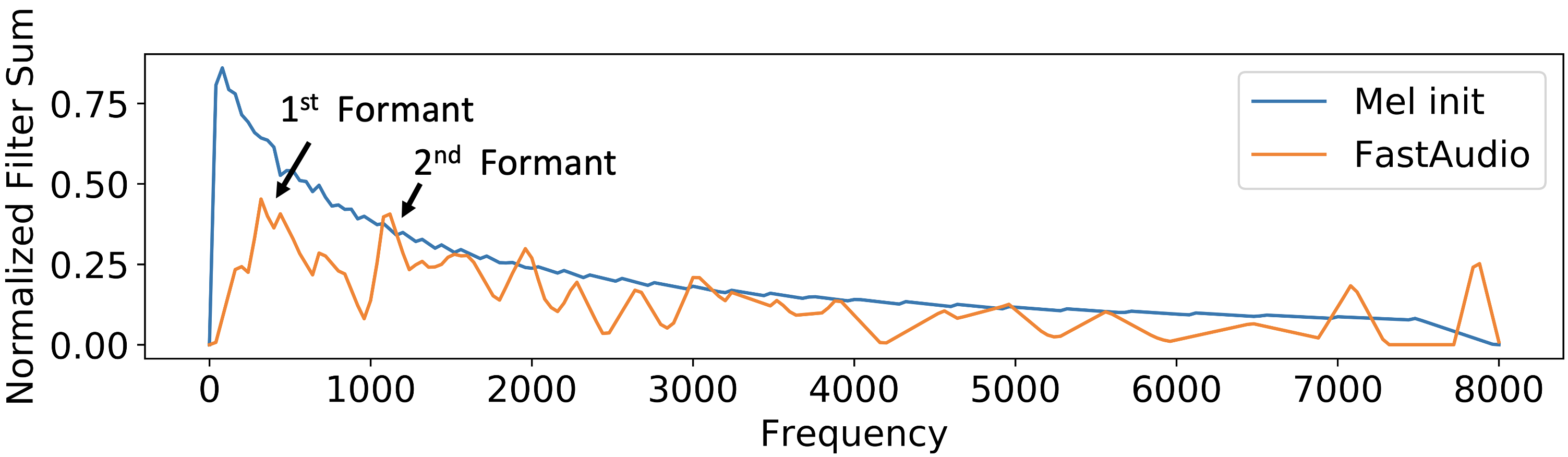}
\caption{Cumulative frequency response of the FastAudio filters}
\label{fig:cumulative}
\end{figure}

Formants are the spectral peaks resulted from acoustic resonance of the human vocal tract. Since in English vowels contain more energy than consonants, we expect our learned filters center frequencies to concentrate around the formants of average vowels of English\cite{Lindblom1990ExplainingPV}. We plotted the cumulative frequency response of the FastAudio in Figure \ref{fig:cumulative}. We found 2 peaks in the lower frequency and 1 peak in the high frequency. The peaks in frequencies around 320Hz$\sim$440 Hz and 1120Hz may correspond to the $1st$ and $2nd$ formants averaged over all vowels in English\cite{Ravanelli2018SpeakerRF}. This adaptation to human speech suggests FastAudio was able to successfully learn what is important for spoof speech detection tasks. Similar adaptation was also reported in FST-based front-end for speech identification tasks\cite{Ravanelli2018SpeakerRF}.

\begin{table*}[htb!]
\begin{center}
\caption{ \label{table:2021}A stage-wise comparison of different Front-end's performance on ASVspoof 2021 LA dataset}
  \begin{tabular}{ccccccccc}
  \hline
     & &  & \multicolumn{2}{c}{\textbf{ECAPA-TDNN}} & \multicolumn{2}{c}{\textbf{X-vector}} \\ 
    \hline
    \textbf{Front-end} & \textbf{\#Params} & \textbf{Constraint} &  & \textbf{min-tDCF}&  & \textbf{min-tDCF} & \textbf{MACs} & \textbf{Train Time/Epoch} \\ 
    \hline
    CQT & 0 & Fixed &  &  0.3676  &  & 0.3812 & 0 & 10:58 min\\
    Fbanks & 0 & Fixed &  &  \textbf{0.26} &  & 0.2788 & 0 & 10:53 min\\
    FastAudio-Tri & 80 & Shape+Clamp &   & 0.2661  &  & 0.3047 & 0 & 13:02 min\\
    FastAudio-Gauss & 80 & Shape+Clamp &   & \textbf{0.2611}& & 0.3122& 0 & 12:51 min\\
    FastAuido-Sort & 80 & Shape+Clamp+Order &  & 0.388 & & 0.293& 0 & 12:59 min\\
    LEAF & 282 & Shape+Clamp &  & 0.2753&  & 0.2794& 0.01GMac & 34.45 min\\
    nnAudio &8.04k & No &   & 0.2783&  & 0.3376 & 0 & 13:00 min\\
    TD-filterbanks & 31k & Shape+Clamp& &  \textbf{0.2522}& & 0.2827& 1.32GMac & 22.48 min\\
    \hline
    \hline
    \textbf{Front-end} & \textbf{Name} & \textbf{Constraint} &  & \textbf{min-tDCF}& &\textbf{EER} &\textbf{Backend} &  \textbf{Baseline}\\ 
    \hline
    SincNet & RawNet2 & Fixed &  &   0.4152& & 9.49& ResNet &  \checkmark \\
    LFCC & LFCC-LCNN & Fixed &  &   0.3152 & & 8.90& LCNN& \checkmark \\
    LFCC & LFCC-GMM & Fixed &  &  0.5836& &21.13& GMM&   \checkmark \\
    CQCC & CQCC-GMM & Fixed &  &  0.4948& &15.80& GMM&   \checkmark \\
    \hline
  \end{tabular}
 \end{center}
\end{table*} 

Interestingly, we also found peaks in the high pitch regions near the sampling boundary, which suggests spoof speech may differ from the real speech in frequencies that are ``ignored'' by scales used by handcrafted front-ends like the Mel-scale. High-frequency energy was thought to be less important and subsequently underrepresented in Mel-scales. However, in the spoof detection task, we suspect that because these high frequencies are ``unimportant'' to human hearing, the spoof speech generator does not create realistic imitation in high frequencies. Thus, the representation of high-frequency data may be a good indicator for spoof speech detection. 

Together, these findings indicated that:
\begin{enumerate}

        \item Learned FastAudio filters are more selective than their initialization.
        \item FastAudio emphasizes frequencies around 1st and 2nd formants, which may be important for distinguishing between spoof and \textit{bona fide} speech.
        \item Learned FastAudio filters are more sensitive to high-frequency energy, which may be a salient feature of spoof detection.
        \item Through end-to-end training, FastAudio can adapt to spoof detection tasks. The front-end successfully adapted to the downstream task and was able to learn the phonetics of human speech.

\end{enumerate}

\subsection{\textbf{How can people use FastAudio for spoof detection and suggestion for model fusion}}

People who focus on designing the back-end for spoof speech detection can use FastAudio as a drop-in replacement. Our experiment shows that FastAudio is a better front-end than CQT in spoof detection, despite CQT being reported as the best front-end in previous research\cite{Li2021ReplayAS}. From the information theory's perspective, the fusion of the result from models whose front-end outputs are least similar tends to produce a better result. Thus, the visualizations of learnable front-ends output in this paper can provide guidance for feature selection, which can be used as a supplement to handcrafted front-ends for spoof detection\cite{Xiao2015SpoofingSD}.

\subsection{\textbf{Does the lesson learned generalize to ASVspoof 2021 dataset?}}

ASVspoof 2021 dataset kept the train and development part of 2019 data the same and added a much larger evaluation dataset that is 10 times the size as before. Thus, the ASVspoof 2021 challenge is more difficult because more spoofing techniques are added. Since the ASVspoof 2021 dataset was not officially released yet, we can not analyze the details as we did on the ASVspoof 2019 dataset. However, since we participated in the 2021 competition, we summarized a preliminary result of different front-end's performances in Table \ref{table:2021}. We found that CQT performs poorly on the new dataset and TD-filterbanks is the best learnable front-end. FastAudio has a similar performance as its hand-crafted equivalent (FBanks).

\section{Conclusion}
\label{conclusionfuturework.section}

This paper investigates the performance of learnable front-ends on spoof detection and proposes an STFT-based audio front-end called FastAudio. We tested the proposed front-end under different constraint settings and found FastAudio was able to successfully adapt to spoof detection. The proposed front-end achieves top performance on the ASVspoof 2019 dataset, beating the fixed equivalent by 27\% and surpassing the performance of CQT, which was reported as the best hand-crafted feature for spoof speech detection. From our work on learnable front-ends for spoof speech detection, we learned the following important lessons:

\begin{enumerate}
    \item Learnable front-end can beat the best handcrafted features in spoof speech detection since they can adapt to the downstream tasks.
    \item Shape constraint is important for filterbank learning to prevent over-fitting.
    \item The center frequencies and bandwidth of filters do not need to be sorted, and
    \item High-frequency information can be important for spoof speech detection.
\end{enumerate}

In future work, we plan to test FastAudio's potential outside spoof detection domain, on other datasets such as UrbanSound8K.

\bibliographystyle{fastaudio}
\bibliography{fastaudio}

\begin{thebibliography}{10}
\providecommand{\url}[1]{#1}
\csname url@samestyle\endcsname
\providecommand{\newblock}{\relax}
\providecommand{\bibinfo}[2]{#2}
\providecommand{\BIBentrySTDinterwordspacing}{\spaceskip=0pt\relax}
\providecommand{\BIBentryALTinterwordstretchfactor}{4}
\providecommand{\BIBentryALTinterwordspacing}{\spaceskip=\fontdimen2\font plus
\BIBentryALTinterwordstretchfactor\fontdimen3\font minus
  \fontdimen4\font\relax}
\providecommand{\BIBforeignlanguage}[2]{{%
\expandafter\ifx\csname l@#1\endcsname\relax
\typeout{** WARNING: IEEEtran.bst: No hyphenation pattern has been}%
\typeout{** loaded for the language `#1'. Using the pattern for}%
\typeout{** the default language instead.}%
\else
\language=\csname l@#1\endcsname
\fi
#2}}
\providecommand{\BIBdecl}{\relax}
\BIBdecl

\bibitem{Bret2021USsmart}
\BIBentryALTinterwordspacing
B.~Kinsella, ``U.s. smart speaker growth flat lined in 2020,'' 2021. [Online].
  Available:
  \url{https://voicebot.ai/2021/04/14/u-s-smart-speaker-growth-flat-lined-in-2020/}
\BIBentrySTDinterwordspacing

\bibitem{Shen2018NaturalTS}
J.~Shen, R.~Pang, R.~J. Weiss, M.~Schuster, N.~Jaitly, Z.~Yang, Z.~Chen,
  Y.~Zhang, Y.~Wang, R.~Skerry-Ryan, R.~Saurous, Y.~Agiomyrgiannakis, and
  Y.~Wu, ``Natural tts synthesis by conditioning wavenet on mel spectrogram
  predictions,'' \emph{2018 IEEE International Conference on Acoustics, Speech
  and Signal Processing (ICASSP)}, pp. 4779--4783, 2018.

\bibitem{Zeghidour2021LEAFAL}
N.~Zeghidour, O.~Teboul, F.~D.~C. Quitry, and M.~Tagliasacchi, ``Leaf: A
  learnable frontend for audio classification,'' \emph{ArXiv}, vol.
  abs/2101.08596, 2021.

\bibitem{Todisco2019ASVspoof2F}
M.~Todisco, X.~Wang, V.~Vestman, M.~Sahidullah, H.~Delgado, A.~Nautsch,
  J.~Yamagishi, N.~Evans, T.~Kinnunen, and K.-A. Lee, ``Asvspoof 2019: Future
  horizons in spoofed and fake audio detection,'' \emph{ArXiv}, vol.
  abs/1904.05441, 2019.

\bibitem{Brown1991CalculationOA}
J.~C. Brown, ``Calculation of a constant q spectral transform,'' \emph{Journal
  of the Acoustical Society of America}, vol.~89, pp. 425--434, 1991.

\bibitem{Li2021ReplayAS}
X.~Li, N.~Li, C.~Weng, X.~Liu, D.~Su, D.~Yu, and H.~Meng, ``Replay and
  synthetic speech detection with res2net architecture,'' in \emph{ICASSP},
  2021.

\bibitem{Andn2014DeepSS}
J.~And{\'e}n and S.~Mallat, ``Deep scattering spectrum,'' \emph{IEEE
  Transactions on Signal Processing}, vol.~62, pp. 4114--4128, 2014.

\bibitem{Lee2018SampleCNNED}
J.~Lee, J.~Park, K.~L. Kim, and J.~Nam, ``Samplecnn: End-to-end deep
  convolutional neural networks using very small filters for music
  classification,'' \emph{Applied Sciences}, vol.~8, p. 150, 2018.

\bibitem{Desplanques2020ECAPATDNNEC}
B.~Desplanques, J.~Thienpondt, and K.~Demuynck, ``Ecapa-tdnn: Emphasized
  channel attention, propagation and aggregation in tdnn based speaker
  verification,'' in \emph{INTERSPEECH}, 2020.

\bibitem{Villalba2020StateoftheartSR}
J.~Villalba, N.~Chen, D.~Snyder, D.~Garcia-Romero, A.~McCree, G.~Sell,
  J.~Borgstrom, L.~P. Garc{\'i}a-Perera, F.~Richardson, R.~Dehak,
  P.~Torres-Carrasquillo, and N.~Dehak, ``State-of-the-art speaker recognition
  with neural network embeddings in nist sre18 and speakers in the wild
  evaluations,'' \emph{Comput. Speech Lang.}, vol.~60, 2020.

\bibitem{Lippmann1997SpeechRB}
R.~Lippmann, ``Speech recognition by machines and humans,'' \emph{Speech
  Commun.}, vol.~22, pp. 1--15, 1997.

\bibitem{Zwicker1980AnalyticalEF}
E.~Zwicker and E.~Terhardt, ``Analytical expressions for critical‐band rate
  and critical bandwidth as a function of frequency,'' \emph{Journal of the
  Acoustical Society of America}, vol.~68, pp. 1523--1525, 1980.

\bibitem{Glasberg1990DerivationOA}
B.~Glasberg and B.~Moore, ``Derivation of auditory filter shapes from
  notched-noise data,'' \emph{Hearing Research}, vol.~47, pp. 103--138, 1990.

\bibitem{Ravanelli2018SpeakerRF}
M.~Ravanelli and Y.~Bengio, ``Speaker recognition from raw waveform with
  sincnet,'' \emph{2018 IEEE Spoken Language Technology Workshop (SLT)}, pp.
  1021--1028, 2018.

\bibitem{Zeghidour2018LearningFF}
N.~Zeghidour, N.~Usunier, I.~Kokkinos, T.~Schatz, G.~Synnaeve, and E.~Dupoux,
  ``Learning filterbanks from raw speech for phone recognition,'' \emph{2018
  IEEE International Conference on Acoustics, Speech and Signal Processing
  (ICASSP)}, pp. 5509--5513, 2018.

\bibitem{Baevski2020wav2vec2A}
A.~Baevski, H.~Zhou, A.~rahman Mohamed, and M.~Auli, ``wav2vec 2.0: A framework
  for self-supervised learning of speech representations,'' \emph{ArXiv}, vol.
  abs/2006.11477, 2020.

\bibitem{Tak2021EndtoEndAW}
H.~Tak, J.~Patino, M.~Todisco, A.~Nautsch, N.~Evans, and A.~Larcher,
  ``End-to-end anti-spoofing with rawnet2,'' in \emph{ICASSP}, 2021.

\bibitem{Cheuk2020nnAudioAO}
K.~Cheuk, H.~Anderson, K.~R. Agres, and D.~Herremans, ``nnaudio: An on-the-fly
  gpu audio to spectrogram conversion toolbox using 1d convolutional neural
  networks,'' \emph{IEEE Access}, vol.~8, pp. 161\,981--162\,003, 2020.

\bibitem{Sainath2013LearningFB}
T.~Sainath, B.~Kingsbury, A.~rahman Mohamed, and B.~Ramabhadran, ``Learning
  filter banks within a deep neural network framework,'' \emph{2013 IEEE
  Workshop on Automatic Speech Recognition and Understanding}, pp. 297--302,
  2013.

\bibitem{Yu2017DNNFB}
H.~Yu, Z.~Tan, Y.~Zhang, Z.~Ma, and J.~Guo, ``Dnn filter bank cepstral
  coefficients for spoofing detection,'' \emph{IEEE Access}, vol.~5, pp.
  4779--4787, 2017.

\bibitem{Zhang2019DiscriminativeFF}
T.~Zhang and J.~Wu, ``Discriminative frequency filter banks learning with
  neural networks,'' \emph{EURASIP Journal on Audio, Speech, and Music
  Processing}, vol. 2019, pp. 1--16, 2019.

\bibitem{snyder2018x}
D.~Snyder, D.~Garcia-Romero, G.~Sell, D.~Povey, and S.~Khudanpur, ``X-vectors:
  Robust dnn embeddings for speaker recognition,'' in \emph{2018 IEEE
  International Conference on Acoustics, Speech and Signal Processing
  (ICASSP)}.\hskip 1em plus 0.5em minus 0.4em\relax IEEE, 2018, pp. 5329--5333.

\bibitem{Ravanelli2021SpeechBrainAG}
M.~Ravanelli, T.~Parcollet, P.~Plantinga, A.~Rouhe, S.~Cornell, L.~Lugosch,
  C.~Subakan, N.~Dawalatabad, A.~Heba, J.~Zhong, J.-C. Chou, S.-L. Yeh, S.-W.
  Fu, C.-F. Liao, E.~Rastorgueva, F.~Grondin, W.~Aris, H.~Na, Y.~Gao, R.~Mori,
  and Y.~Bengio, ``Speechbrain: A general-purpose speech toolkit,''
  \emph{ArXiv}, vol. abs/2106.04624, 2021.

\bibitem{desplanques2020ecapa}
B.~Desplanques, J.~Thienpondt, and K.~Demuynck, ``Ecapa-tdnn: Emphasized
  channel attention, propagation and aggregation in tdnn based speaker
  verification,'' \emph{arXiv preprint arXiv:2005.07143}, 2020.

\bibitem{Xiao2015SpoofingSD}
X.~Xiao, X.~Tian, S.~Du, H.~Xu, C.~E. Siong, and H.~Li, ``Spoofing speech
  detection using high dimensional magnitude and phase features: the ntu
  approach for asvspoof 2015 challenge,'' in \emph{INTERSPEECH}, 2015.

\bibitem{Lindblom1990ExplainingPV}
B.~Lindblom, ``Explaining phonetic variation: A sketch of the h\&h theory,''
  1990.

\end{thebibliography}

\end{document}